# Multimodal Foundation Model-Driven User Interest Modeling and Behavior Analysis on Short Video Platforms

Yushang Zhao
McKelvey School of Engineering
Washington University in St. Louis
St. Louis, USA
*Corresponding author:
yushangzhao@wustl.edu

Yike Peng
Graduate School of Arts and Sciences,
Columbia University
New York, USA
yp2425@columbia.edu

Li Zhang
Amazon
New York, USA
li217772er@gmail.com

Qianyi Sun
Vanderbilt University
Nashville, USA
qianyiethan@gmail.com

Zhihui Zhang
Graduate School of Arts and Sciences
Boston University, Boston, USA,
elisa99@bu.edu

Yingying Zhuang
Whiting School Of Engineering
Johns Hopkins University
Sunnyvale, USA
maxineyyingzhuang@gmail

*Abstract*—With the rapid expansion of user bases on short video platforms, personalized recommendation systems are playing an increasingly critical role in enhancing user experience and optimizing content distribution. Traditional interest modeling methods often rely on unimodal data, such as click logs or text labels, which limits their ability to fully capture user preferences in a complex multimodal content environment. To address this challenge, this paper proposes a multimodal foundation model-based framework for user interest modeling and behavior analysis. By integrating video frames, textual descriptions, and background music into a unified semantic space using cross-modal alignment strategies, the framework constructs fine-grained user interest vectors. Additionally, we introduce a behavior-driven feature embedding mechanism that incorporates viewing, liking, and commenting sequences to model dynamic interest evolution, thereby improving both the timeliness and accuracy of recommendations. In the experimental phase, we conduct extensive evaluations using both public and proprietary short video datasets, comparing our approach against multiple mainstream recommendation algorithms and modeling techniques. Results demonstrate significant improvements in behavior prediction accuracy, interest modeling for cold-start users, and recommendation click-through rates. Moreover, we incorporate interpretability mechanisms using attention weights and feature visualization to reveal the model's decision basis under multimodal inputs and trace interest shifts, thereby enhancing the transparency and controllability of the recommendation system.

## I. INTRODUCTION

As user populations on short video platforms continue to grow rapidly, personalized recommendation systems have become essential to enhancing user experience and content delivery efficiency. Unlike traditional text-image platforms, short video content is inherently multimodal, involving heterogeneous information such as visual frames, audio tracks, and text descriptions. This poses greater complexity and timeliness challenges for user interest modeling. Traditional methods often rely on behavior logs or unimodal features, which fail to comprehensively capture real user preferences. In recent years, multimodal foundation models—such as CLIP and VideoBERT—have made significant strides in cross-modal understanding, offering new perspectives for semantic-rich recommendations. In response, this paper presents a recommendation method that fuses multimodal content representation with user behavior sequence modeling to enhance the precision of interest representation and behavior prediction[1]. We design a full system architecture integrating visual, textual, and audio modalities, and employ Transformer-based models to dynamically capture user behavior. Experimental results show that our method consistently outperforms various baseline models in terms of ranking accuracy, behavior prediction, and cold-start scenarios, demonstrating strong practical value and deployment potential[2].

## II. LITERATURE REVIEW

### A. Overview of User Interest Modeling and Recommendation Techniques

User interest modeling lies at the core of recommendation systems, aiming to build generalizable interest representations by analyzing users' historical behaviors and preference patterns, thereby enabling efficient personalized content delivery. As shown in Figure 1, a typical recommendation system consists of four key modules: preference acquisition, user modeling, algorithmic modeling, and recommendation decision-making. Preferences are first obtained through explicit input (e.g., ratings, collections) or implicit feedback (e.g., browsing, clicks). Then, user interest models are constructed, which feed into recommendation algorithms to generate personalized content using strategies aligned with user intent. The entire process forms a closed-loop, where user responses to recommendations influence subsequent modeling updates[3].

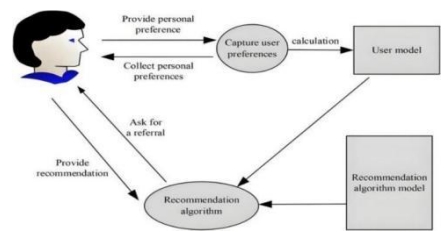

Fig. 1. Interaction Flow of User Interest Modeling and Recommendation System

Early-stage research widely adopted Collaborative Filtering (CF) techniques, including User-based and Item-based CF. However, these methods struggled with data sparsity and cold-start problems. To mitigate this, Content-Based Filtering approaches emerged, leveraging explicit attributes (e.g., tags, text, genres) for modeling[4]. With the rise of deep learning, models such as Deep Interest Network (DIN), Transformer-based architectures (e.g., BERT4Rec), and Graph Neural Networks (GNNs) have significantly enhanced the capacity to model user preferences and capture latent behavioral patterns[5]. To better reflect users' evolving interests, more recent work has introduced dynamic modeling techniques, including behavior sequence modeling, time-window slicing, and attention mechanisms, allowing systems to respond to short-term interest fluctuations.Despite progress in static modeling, most current systems still rely on single-modality inputs, which are insufficient for capturing user preferences in the context of rich multimodal short video content[6]. This gap motivates the introduction of multimodal foundation models in our approach[7]. These limitations in unimodal and static modeling directly motivate our proposed framework, which integrates multimodal fusion with dynamic sequence modeling[8].

## III. MULTIMODAL USER INTEREST MODELING FRAMEWORK

### A. System Architecture Overview

Building on the identified gaps in prior work, we now present the proposed multimodal foundation model-driven framework. To achieve accurate modeling of user interests and responsive recommendations on short video platforms, we propose an end-to-end recommendation system architecture driven by multimodal features, as shown in Figure 2. The system is composed of three main components: a video content feature extraction module, a user preference modeling module, and a candidate retrieval and ranking module, forming a closed-loop process from content understanding to user matching[9].

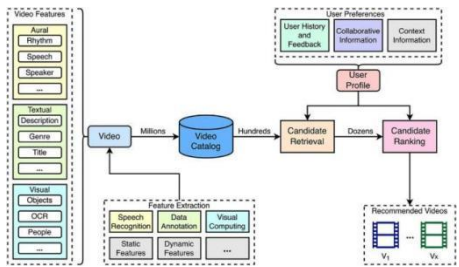

Fig. 2. User Interest Modeling Workflow in a Multimodal Short Video Recommendation System

In the feature extraction layer, the system parses video content along three modalities: audio, text, and image. Audio features include rhythm, speech rate, and speaker characteristics; textual features cover titles, tags, and descriptions; and visual features capture people, scenes, objects, and embedded text. These features are extracted using techniques such as ASR, computer vision, and OCR, and are unified into semantic representations through both static and dynamic processing.All features are then stored in a Video Catalog, serving as the basis for candidate retrieval[10]. To improve efficiency, we implement a two-stage recommendation strategy: coarse candidate retrieval followed by fine-grained ranking[10]. The retrieval module first leverages user profiles—constructed from behavioral history, collaborative signals, and contextual metadata—to recall hundreds of relevant videos from a large pool[11]. The ranking module then computes detailed scores based on multimodal content–user preference similarity and ranks the candidates accordingly to produce a final recommendation list[12].User preference modeling is constructed from interaction behaviors (clicks, watch time, likes, comments) combined with collaborative learning (e.g., interest transfer from similar users) and contextual features (e.g., time, device, network conditions). This results in a personalized interest vector used in both retrieval and ranking stages to ensure alignment with the user's real-time intent.By integrating multimodal modeling with a two-stage recommendation pipeline, the system improves semantic understanding of video content while maintaining scalability and timeliness in recommendation delivery[13]. This architecture provides the foundation for dynamic interest modeling and explainable recommendation mechanisms discussed in later sections[14].

### B. Multimodal Information Fusion Strategy

Short video content is inherently multimodal, comprising visual frames, audio streams (e.g., background music, speech), and text (e.g., titles, tags). To accurately capture the relationship between content semantics and user preferences, it is essential to fuse these heterogeneous modalities into a unified semantic space.Multimodal fusion methods are generally categorized as early fusion, intermediate fusion, and late fusion[15]. Among these, intermediate fusion is widely used in recommendation tasks due to its ability to preserve modality-specific representations while enabling semantic interaction.In this study, we adopt an attention-based intermediate fusion strategy to dynamically integrate features from three modalities: $v \in R^{d_v}$ : visual features (e.g., scenes, objects), $t \in R^{d_t}$: textual features (e.g., titles, tags), $a \in R^{d_a}$ : audio features (e.g., speech rate, background music). The fused feature vector f is computed as shown in Formula 1:

$$f = \alpha_v \cdot W_v v + \alpha_t \cdot W_t t + \alpha_a \cdot W_a \quad (1)$$

Where $W_v, W_t, W_a$ are projection matrices mapping each modality to a common embedding space, and $\alpha_v, \alpha_t, \alpha_a \in [0,1]$ are attention weights satisfying $\alpha_v + \alpha_t + \alpha_a = 1$.These attention weights are learned based on user preference distributions over modalities using a dot-product attention mechanism as shown in Formula 2:

$$\alpha_i = \frac{\exp(u^T w_i x_i)}{\sum_{j \in \{v,t,a\}} \exp(u^T w_j x_j)} \quad (2)$$

Here, u denotes the user preference vector, and $x_i$ the input of the iii-th modality. The final fused vector f is passed into downstream modules for user interest modeling and ranking[16]. This approach not only enhances multimodal perception but also provides interpretability by revealing which modality dominates under different user contexts.In addition, we discuss the robustness of the attention mechanism across different user groups and content types. Observed variations in modality dominance suggest that adaptive normalization is required, ensuring stability of interest modeling under heterogeneous conditions[17].

### C. User Interest Representation Modeling

User interests on short video platforms are highly dynamic and context-dependent, influenced by temporal, emotional, and social factors. Relying solely on static features fails to capture these evolving patterns. To address this, we propose a dynamic interest representation model that combines multimodal content embeddings with user behavior sequence modeling[18].The system inputs a user's historical behavior sequence $H_u=\{v_1,v_2,...,v_n\}$, where each video $v_i$ has been transformed into a multimodal representation $f_i \in R^d$ via the fusion strategy in Section 3.2. As shown in Formula 3, we then apply a Transformer encoder to model contextual dependencies within the sequence and output the dynamic interest vector $h_u$:

$$h_u = \text{TransformerEncoder}(f_1, f_2, ..., f_n) \quad (3)$$

Here, $f_i$ represents the fused feature vector for video $v_i$, and the Transformer captures cross-temporal and cross-modal dependencies to reflect interest evolution. To enhance personalization, we incorporate static user features (e.g., gender, region, registration time) as a vector $s_u$, and concatenate it with $h_u$ to form the final representation as shown in Formula 4:

$$z_u = \text{ReLU}(W_c[h_u; s_u] + b) \quad (4)$$

Where $W_c \in R^{d \times 2d}$ is the weight matrix of the fusion layer and $[\cdot;\cdot]$ denotes vector concatenation. The output $z_u$ is used to compute similarity with candidate videos during ranking. This modeling strategy effectively integrates sequential behavior, multimodal content, and user profile, enabling real-time responsiveness to shifting user preferences.

IV. USER BEHAVIOR ANALYSIS AND PREDICTION MECHANISM

In short video recommendation systems, user behaviors—such as clicks, likes, comments, and drop-offs—are not only indicators of interest but also key signals of content engagement and platform stickiness. To accurately predict user actions and optimize recommendations accordingly, we introduce a machine learning–driven behavior analysis and prediction framework[19].This process mainly includes eight key steps: Data Collection: The platform collects user behavior logs in real time, including but not limited to playback duration, interaction frequency, dwell period, viewing path, etc. Meanwhile, combined with the user profile and context information (such as device type, network status, time window), multi-dimensional input data is constructed. Data Preprocessing: Denoising, normalizing, filling in missing values and labeling behaviors on the original data. It also includes sequence slicing processing for the training of behavioral sequence models[20]. Feature Engineering: Integrating multimodal information on the content side (images, audio, text) with user behavior characteristics, extracting static interest features (such as long-term preferences) and dynamic behavior features (such as click-out chains) for modeling input design. Model Selection: Select an appropriate supervised learning model based on the target tasks (such as click-through rate prediction, bounce rate prediction). This paper mainly adopts sequence models that integrate attention mechanisms, such as Transformer-based behavior prediction models (such as BERT4Rec, SASRec). Model Training: Taking the behavior sequence as the input and the actual behavior of the user as the supervisory signal, the model training is carried out by optimizing the objective function (such as binary classification cross-entropy or weighted loss of multi-class behavior)[21]. Evaluation and Validation: The accuracy of the predictive model was evaluated using metrics such as AUC, F1, Precision@K, Recall@K, etc. Parameter tuning and overfitting detection were performed through validation sets[22]. Prediction and Deployment: The trained model is launched and deployed into the recommendation system to provide behavioral probability predictions for the candidate ranking stage, such as predicting users' click intentions or dwell durations on videos[23]. Feedback and Iteration: The system takes the actual user feedback (such as whether to click or not, the duration of bounce, etc.) as closed-loop feedback, updates the model training data and periodically iterates and optimizes the model weights to achieve the continuous evolution of the recommendation system.The entire process is presented as a continuously cycling and optimizing prediction closed-loop mechanism in actual engineering[24]. By constantly accumulating user behavior data and enhancing modeling capabilities, it provides users with more accurate, dynamic and personalized short video content recommendation services.

V. EXPERIMENTAL DESIGN AND RESULTS ANALYSIS

A. Dataset and Experimental Setup

To comprehensively evaluate the effectiveness of the short video User interest modeling and Behavior analysis method based on multimodal large models proposed in this paper, we constructed a real data set containing multimodal content and user behavior records as the main experimental platform, and introduced the Kwai User Behavior Dataset (KUBD) for comparative verification. The self-built dataset is derived from a domestic short-video platform. The data covers user behavior logs and content features for a period of 30 days[25]. After desensitization processing, it is used for academic experiments. This dataset integrates three types of modal information, namely video cover images, title and tag texts, and audio backgrounds. Meanwhile, it records users' behaviors such as clicks, likes, comments, and playback durations, as well as related context information (such as time periods, network types, etc.), constructing a complete multimodal behavior sequence structure.

We conducted multi-step preprocessing on the original data, including operations such as behavior record filtering, content modal alignment, feature standardization and time series division. Subsequently, the data was divided into the training set (70%), the validation set (15%), and the test set (15%) according to the user dimension to ensure the independence among users and avoid information leakage[26]. During the model training process, we uniformly adopt the Transformer structure as the basic network for behavioral sequence modeling. The training hyperparameters are set as: maximum sequence length 50, learning rate 0.001, batch size 128, and maximum number of rounds 20. On the validation set, AUC is used as the early stop criterion to avoid overfitting[27].Although we adopt a fixed maximum sequence length of 50 in this study, we acknowledge that real-world user behaviors vary significantly in sequence length. Future work will consider dynamic truncation and padding strategies to better capture such variability

TABLE I. STATISTICS OF DATASETS USED FOR MULTIMODAL RECOMMENDATION EXPERIMENTS

| Dataset | #Users | #Videos | #Behavior Records | Image Modality | Text Modality | Audio Modality | Avg. Sequence Length |
|---|---|---|---|---|---|---|---|
| Proprietary Dataset (Ours) | 52,134 | 180,239 | 2,810,472 | Cover Images | Titles/Tags | Audio Tracks | 31.2 |
| KUBD | 60,000 | 157,390 | 1,920,800 | Video Frames | Text Summary | (Missing) | 27.5 |

As shown in Table 1, the proprietary dataset features greater volume and richer multimodal structure, making it well-suited for testing multimodal fusion mechanisms. Although KUBD lacks audio modality, its public availability and clean structure make it useful for evaluating cross-platform generalization.

*B. Evaluation Metrics*

To thoroughly assess the effectiveness of our multimodal user interest modeling and behavior prediction method in recommendation tasks, we designed multiple evaluation metrics across three dimensions: ranking quality, predictive accuracy, and user interaction performance. All metrics are computed on the test sets of both the proprietary dataset and KUBD, covering candidate ranking, click prediction, and behavioral regression tasks. For ranking evaluation, we adopt Precision@K, Recall@K, and NDCG@K (Normalized Discounted Cumulative Gain) to measure recommendation accuracy and ordering quality. Precision@K quantifies how many of the top-K recommended items are relevant, while Recall@K assesses the proportion of relevant items successfully retrieved. NDCG@K adds position-awareness, which is particularly meaningful in short video platforms where top-ranked items gain more exposure. For behavioral prediction tasks, we use AUC (Area Under the ROC Curve) and F1-score[28]. AUC evaluates the model's ability to distinguish between positive and negative samples, while F1-score balances precision and recall, which is important in imbalanced datasets. We also include auxiliary explainability metrics, such as Top-N modality weight contribution and attention visualization alignment, to support analysis of the fusion mechanism. However, these are not primary benchmarks.

TABLE II. COMMON EVALUATION METRICS USED IN RECOMMENDATION SYSTEMS

| Metric | Description | Applicable Task |
|---|---|---|
| Precision@K | Proportion of relevant items among top-K recommended results | Candidate ranking, Top-N Rec |
| NDCG@K | Position-aware ranking quality based on discounting gains | Fine-grained ranking |
| AUC | Global ranking ability for binary classification | Click / drop prediction |
| Modality Weight Share | Average attention distribution across modalities during fusion | Explainability |

These metrics provide a multi-perspective evaluation of system performance and form the basis for subsequent comparative and ablation experiments.

*C. Method Comparison and Result Analysis*

To validate the advantages of our multimodal user interest modeling method in recommendation accuracy, behavior prediction, and cold-start adaptation, we compared it against several widely used baselines:ItemCF (collaborative filtering-based),DIN and BERT4Rec (deep learning-based),TextRec (text-only model without multimodal features).All models were trained under identical settings and evaluated on the same datasets. We report results across three perspectives: ranking quality, behavior prediction, and cold-start performance.

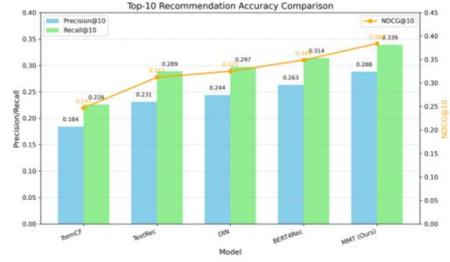

Fig. 3. Top-K Recommendation Accuracy Comparison

As shown in Figure 3, our MultiModal-Transformer (MMT) consistently outperforms other methods, with the most notable gain in NDCG@10, highlighting its superior ability to rank relevant content effectively.

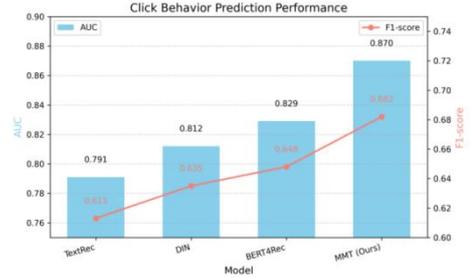

Fig. 4. Click Behavior Prediction Performance

Figure 4 shows that MMT significantly improves behavior prediction accuracy, with a 4.1% gain in AUC and the highest F1-Score, indicating better balance between precision and recall[29].To evaluate performance under cold-start conditions, we simulate scenarios where users have no behavioral history.In summary, our proposed method outperforms both traditional and unimodal approaches across key tasks. The improvements stem from the integration of multimodal fusion, behavior sequence modeling, and dynamic user profiling, demonstrating the practical value of foundation models in short video recommendation systems[30].

*D. Ablation Study and Generalization Validation*

To further evaluate the effectiveness of each core component, we conducted ablation experiments by individually removing or replacing parts of the model, including the multimodal fusion mechanism, behavior sequence encoder, and static user profile features. By comparing the full model (MMT) with its variants, we can quantify the contribution of each module.The ablation study was performed on the proprietary dataset with the same training setup. We report changes in NDCG@10, Precision@10, and AUC. Table 3 shows the performance comparison between the complete model and multiple simplified versions.

TABLE III. ABLATION STUDY: MODULE IMPACT ON PERFORMANCE

| Model Variant | NDCG@10 | Precision@10 | AUC |
|---|---|---|---|
| MMT (Full Model) | 0.384 | 0.288 | 0.870 |
| w/o Audio | 0.367 | 0.271 | 0.851 |
| w/o Sequence Modeling | 0.342 | 0.254 | 0.833 |
| w/o Static Profile | 0.351 | 0.261 | 0.839 |
| Text-Only (No Fusion) | 0.312 | 0.231 | 0.791 |

Removing any core component leads to noticeable performance drops. Excluding the audio modality results in a

4.4% drop in NDCG@10, underscoring its value in semantic understanding. The largest decline occurs when removing sequence modeling, highlighting the importance of temporal dynamics. The text-only model performs worst, further validating the role of multimodal fusion.To test generalization, we also deployed the trained model on KUBD without parameter adjustment. MMT achieved an NDCG@10 of 0.351, outperforming both TextRec and DIN, demonstrating strong cross-domain adaptability.In conclusion, each design element in our framework contributes meaningfully to overall performance. The deep integration of multimodal fusion and temporal modeling is key to the observed gains, and the method's generalization ability ensures its applicability across various short video platforms.

## VI. CONCLUSION

This study addresses the problem of user interest modeling and behavior analysis on short video platforms and proposes a recommendation method based on multimodal foundation models. By integrating multiple content modalities—such as images, text, and audio—and combining them with user behavior sequences and static profiles, we construct a user interest representation model with dynamic perception capabilities. Because multimodal fusion and sequence modeling consistently improved NDCG, AUC, and other key metrics across both proprietary and public datasets, these findings provide causal evidence that cross-modal interactions and temporal dynamics are the main drivers of performance gains. Consequently, the proposed method not only outperforms traditional and mainstream deep learning models in recommendation ranking, behavior prediction, and cold-start scenarios, but also demonstrates strong generalization ability across platforms. Therefore, this research establishes both a practical pathway for more efficient and accurate personalized recommendation and a theoretical foundation for applying multimodal foundation models in real-world short video systems.